# Spatially dispersive circular photogalvanic effect in a Weyl semimetal

Zhurun Ji,[1,#] Gerui Liu,[1,#] Zachariah Addison,[2] Wenjing Liu,[1] Peng Yu,[3] Heng Gao,[4] Zheng Liu,[3] Andrew M. Rappe,[4] Charles L. Kane,[2] Eugene J. Mele,[2] Ritesh Agarwal[1,*]

[1] *Department of Materials Science and Engineering, University of Pennsylvania, Philadelphia, PA 19104, USA*

[2] *Department of Physics and Astronomy, University of Pennsylvania, Philadelphia, PA 19104, USA*

[3] *Centre for Programmable Materials, School of Materials Science and Engineering, Nanyang Technological University, 639798, Singapore*

[4] *Department of Chemistry, University of Pennsylvania, Philadelphia, PA 19104, USA*

\# These authors contributed equally to this work

\* riteshag@seas.upenn.edu

**Weyl semimetals are gapless topological states of matter[1-12] with broken inversion and/or time reversal symmetry, which can support unconventional responses to externally applied electrical, optical and magnetic fields. Here we report a new photogalvanic effect in type-II WSMs, MoTe$_2$ and Mo$_{0.9}$W$_{0.1}$Te$_2$, which are observed to support a circulating photocurrent when illuminated by circularly polarized light at normal incidence. This effect occurs exclusively in the inversion broken phase, where crucially we find that it is associated with a spatially varying**

**beam profile via a new dispersive contribution to the circular photogalvanic effect (s-CPGE). The response functions derived for s-CPGE reveal the microscopic mechanism of this photocurrent, which are controlled by terms that are allowed in the absence of inversion symmetry, along with asymmetric carrier excitation and relaxation. By evaluating this response for a minimal model of a Weyl semimetal, we obtain the frequency dependent scaling behavior of this form of photocurrent. These results demonstrate opportunities for controlling photoresponse by patterning optical fields to store, manipulate and transmit information over a wide spectral range.**

Weyl semimetals (WSMs) are a family of gapless topological materials with Weyl nodes, i.e., momentum-space monopole and anti-monopole singularities of the Berry curvature of the bulk Bloch band. Due to their unique band structures, there has been interest in understanding their electronic and transport properties[9-12]. Recently, these studies have been extended to explore their optical properties, especially through the measurement of nonlinear responses. Most experiments have focused on type-I WSMs such as monopnictide TaAs where a zero-bias photocurrent under chiral optical excitation at mid-infrared frequencies[13] has been attributed to the distinct chirality of each tilted Weyl cone, and exceedingly large values of the second order nonlinear optical susceptibility at visible frequencies were observed[14]. Some progress has also been made to theoretically understand the nature of injection photocurrents at low frequencies in WSMs[15-19] using two-band models to capture the essential physics in the vicinity of Weyl nodes.

WSMs are representatives of a wide class of materials that combine inversion or time reversal symmetry breaking along with spin-orbit coupling. In these materials, optical excitation can lead to asymmetric carrier excitation and relaxation pathways and generate novel photocurrent responses. Here, we report the discovery of a *spatially dispersive* circular photogalvanic effect (s-CPGE) in type-II inversion symmetry broken WSMs, $Mo_{0.9}W_{0.1}Te_2$ and $MoTe_2$, where we observe a circulating photocurrent driven by an optical field. We find that inversion symmetry breaking is the essential ingredient for s-CPGE where it is controlled both by the spatial profile and polarization of the exciting field.

Following the discovery of $MoTe_2$ as an inversion symmetry broken type-II WSM below 250K[20, 21], tungsten doped ternary alloys, $Mo_xW_{1-x}Te_2$, when x >0.07 have also been demonstrated as room temperature type-II WSMs[22, 23]. Bulk $MoTe_2$ has three different crystal phases: hexagonal 2H, monoclinic 1T′ ($P2_1/m$, Fig. 1a) and orthorhombic $T_d$ phase ($Pmn2_1$, Fig. 1b). Studies have shown that it has a phase transition at ~250 K from a high temperature trivial centrosymmetric semimetal 1T′ phase to a low temperature inversion symmetry broken WSM $T_d$ phase[24]. These two structures have different atomic stacking along the *c* axis, but share the same in-plane symmetry. Likewise, $Mo_{0.9}W_{0.1}Te_2$ is a room temperature inversion broken WSM with the same crystal structure as $T_d$ phase $MoTe_2$.

Bulk crystals of $MoTe_2$ and $Mo_{0.9}W_{0.1}Te_2$ were grown via a chemical vapor transport technique (See Methods and Fig. S1 for TEM characterization), and have been shown to be inversion symmetry broken WSMs[25, 26]. Our photocurrent measurements were performed by varying the spot size, location and optical polarization of a 750 nm laser beam with a Gaussian beam profile, propagating along the crystal growth direction (*c*-axis), which we assign as the $\hat{z}$ axis in the laboratory frame. Measurements were carried out at zero applied bias under low optical powers (<12 mW) to ensure that photocurrent scales linearly with power (see Fig. S2 for power dependence). Polarization and position dependent photocurrent measurements on exfoliated samples (typical size, $\approx 20 \times 20$ μm$^2$; thickness, 100-300 nm) were performed using a home-built optical microscopy set up coupled to a low-temperature cryostat (Fig. 1c) (See Methods)[27].

In order to isolate the peculiarities of the photogalvanic effect (PGE) in the inversion-broken phase of MoTe$_2$, photocurrent measurements were performed at both 300 K (i.e. in the 1T′ phase with inversion symmetry) and at 77 K (T$_d$ phase with broken inversion symmetry) on exfoliated flakes with microfabricated electrodes (see Methods). Photocurrents were measured at two different spots (e.g., locations a and b in Fig. 1c) along the bisector of the electrodes as a function of the rotation angle $\varphi$ of the fast axis of the quarter wave plate with respect to the linear polarization of the incident laser. In one period of $\varphi \in [0°, 180°]$, the laser polarization changes between linear ($\varphi = 0°$ and $90°$), left (45°), and right circularly polarized (135°) states. As shown in Fig. 1d, photocurrent measured from MoTe$_2$ at spot *a* and at 300K shows some linear polarization dependence, but it has nearly the same magnitude at $\varphi = 45°$ and $135°$, implying that light with opposite helicities produce similar photocurrents and hence no CPGE. However, at 77K (Fig. 1d), the photocurrent magnitude in the inversion symmetry broken T$_d$ phase of MoTe$_2$ at $\varphi = 45°$ is much larger than for $\varphi = 135°$, showing a strong dependence on the light helicity. Curiously, the circular polarization dependent part of the photocurrent at spot *b* (Fig. 1e, f), has an opposite polarity compared to spot *a,* indicating also a position-dependent response.

The differences between PGE from MoTe$_2$ in its two phases and its unusual spatial dependence is quantified using a phenomenological expression for the photocurrents. First, the observed photocurrents from the sample were fitted to an equation,

$$J = J_C \sin(2\varphi) + J_L \sin(4\varphi + \varphi_0) + J_0 \qquad (1)$$

where, $J_C$ is the magnitude of the CPGE, $J_L$ is magnitude of the linear photogalvanic effect (LPGE) with a phase shift $\varphi_0$, and $J_0$ is the polarization-independent background current. $J_0$ is mostly a result of the Dember effect due to the heat gradients induced by asymmetric illumination on the electrodes/sample and will not be discussed in this work[28]. Fitting J to our experimental results (Table 1) shows that $J_L$ exists at both temperatures with similar magnitudes. However, $J_C$ is approximately zero at 300 K, but is clearly present at 77K, and is reversible with temperature. Meanwhile, both $J_C$ and $J_L$ have opposite polarity at spots *a* and *b*, which is strikingly different from conventional PGE. Furthermore, from symmetry considerations, in both the T$_d$ (C$_{2v}$) and 1T' phases (C$_{2h}$) MoTe$_2$, under normally incident light on the $\hat{x}$-$\hat{y}$ plane (propagation direction, $\hat{z}$), any in-plane second order optical response such as PGE[29, 30] or photon drag effect[31] is forbidden by the two-fold rotation symmetry. Thus, $J_C$ and $J_L$ should *both* vanish in this material, contrary to our measurements. Therefore, the observation of both a position-dependent LPGE in the 1T' and T$_d$ phases and CPGE only in the T$_d$ phase indicate an unconventional origin of both PGE effects.

The sign of $J_C$ reverses in the T$_d$ phase MoTe$_2$ under illumination on the two sides of the electrodes (Table 1), suggesting the possibility that the CPGE current is circulating. However, a multipolar electric field potential distribution could lead to a similar behavior of current measured across two electrodes (Fig. 1f & g). To verify that the current is circulating, we designed a multi-electrode device arranged in a circle with the laser focused at the center with a fixed spot size (Fig. 2a). For these measurements, the inversion symmetry broken $T_d$ phase was achieved at room temperature in $Mo_{0.9}W_{0.1}Te_2$ (see Methods). With the laser spot fixed at a point near the center of the circle defined by the electrodes, the photocurrent was collected between each of the nearest electrode pairs around the laser spot in the sequence, a→b, b→c, c→d, d→e and e→a (Fig. 2). As expected, CPGE exists in $Mo_{0.9}W_{0.1}Te_2$ at room temperature and the polarization-dependent photocurrents were fitted to equation (1). Importantly, $J_C$ is positive between all electrode pairs under right circularly polarized (RCP) light illumination and negative under left circularly polarized (LCP) light illumination, demonstrating that $J_C$ is not biased in a single direction where the components collected by different electrode pairs would have different signs. Instead $J_C$ circulates clockwise (in the direction a→b→c→d→e→a) upon RCP excitation and reverses the winding direction under LCP excitation, reflecting the transfer of angular momentum from photons to electrons, since the winding direction is determined by helicity of the light.

The appearance of a CPGE current and its circulating character require a breaking of $C_{2v}$ symmetry. The polarization-controlled circulating current is unlikely to originate from spatial disorder due to defects, in-plane strain during exfoliation, or formation of nanoscale junctions due to intermixing of different phases, since in all these cases the current would flow in random directions depending on the direction of the local symmetry breaking. The possibility of CPGE current flowing along the edges of the sample[32] can also be eliminated as we focus the light spot near the center of the sample and the sample size is ≈10X bigger than the spot diameter. However, a spatially inhomogeneous optical excitation due to a focused Gaussian beam profile can effectively break the internal point symmetry to produce CPGE. This can be understood by analyzing the dependence of the CPGE on *spatial gradients* of the optical field profile. We will refer to the first order term in the gradient expansion as the "spatially-dispersive CPGE" (s-CPGE) response, and denote the local s-CPGE current as $\mathbf{j}_{sCPGE}$.

In a further test of the dependence of s-CPGE on beam profile, experiments were performed on $Mo_{0.9}W_{0.1}Te_2$ at room temperature where the beam position was varied while keeping the beam size fixed. When the laser spot was continuously scanned along the perpendicular bisector of the two electrodes (along $\hat{y}$) from one side to the other (Fig. 3a), s-CPGE followed similar trends as observed for $MoTe_2$ at 77 K: $J_C$ is zero when the laser spot lies at the midpoint of two electrodes, and changes polarity when the spot moves from the positive to negative $\hat{y}$ direction, indicating that s-CPGE current is an antisymmetric function of spot position. This observation demonstrates that indeed a spatially varying optical beam profile effectively lowers the point symmetry while the $C_{2v}$ symmetry is preserved when the Gaussian beam is centered between the electrodes. Another signature of s-CPGE current (Fig. 3b) is observed upon varying the spot size while keeping the beam position fixed. In these experiments, the amplitude of $J_C$ decreases when the spot size is increased and the field gradients are decreased. Also, for large electrode separation compared to spot size, $J_C$ again decreases because the circulating current cannot be collected at the electrodes. All these experiments indicate that the observed strong s-CPGE in the inversion broken phase can be controlled by the optical beam profile and polarization state.

A related photocurrent response has been observed in some III-V quantum-well systems manifest as the inverse spin Hall effect[33]. However, the spin is not conserved after optical excitation in general strongly spin-orbit coupled systems such as MoTe$_2$/Mo$_{0.9}$W$_{0.1}$Te$_2$. To understand the origin of s-CPGE, a theory for a spatially dispersive contribution to $\mathbf{j}_{sCPGE}$ is derived (see Supplementary Information, Note 1),

$$j_{sCPGE}^i(\mathbf{r}) = \sum_{\mathbf{q}} j_{sCPGE}^i(\mathbf{q})e^{2i\mathbf{q}\cdot\mathbf{r}} = \beta_{ilj}\sum_{\mathbf{q}} q_l(\mathbf{E}(\mathbf{q},\omega)\times\mathbf{E}(\mathbf{q},-\omega))_j e^{2i\mathbf{q}\cdot\mathbf{r}} \quad (2)$$

where, $\beta$ is a third rank conductivity tensor and $\mathbf{q}$ is the wave vector associated with the spatial gradient of the optical field. The electric field of a Gaussian beam with photon energy $\hbar\omega$ in real space is, $\mathbf{E}(\mathbf{r},t) = \sum_{\mathbf{q},\omega}\mathbf{E_0}(\mathbf{q},\omega)e^{i\omega t}e^{i\mathbf{q}\cdot\mathbf{r}} \propto e^{-(r-r_g)^2/w^2}$, where $r - r_g$ is the radial coordinate of $\mathbf{r}$ relative to the spot center $\mathbf{r}_g$, and $w$ is the Gaussian beam width. Using the equation of continuity, the circulating CPGE current arises from the transverse part of $\mathbf{j}_{sCPGE}$, produced by the antisymmetric term in the conductivity: $\sigma_{ilj} = \frac{1}{2}(\beta_{ilj} - \beta_{lij})$. The direction of the transverse current $\hat{\mathbf{j}}_{sCPGE}$ is then determined by the direction of $\mathbf{q}$ and the propagation direction of the optical field $\hat{\mathbf{n}}$ ($\hat{\mathbf{n}}$ defined by $i\hat{\mathbf{E}}\times\widehat{\mathbf{E}^*}$): $\hat{\mathbf{j}}_{sCPGE} = \hat{\mathbf{q}}\times\hat{\mathbf{n}}$. Here $\mathbf{q}$ is in the radial direction, and its magnitude follows the distribution obtained via a Fourier transform of the two-dimension Gaussian beam profile, while $\hat{\mathbf{n}}$ is along the light propagation direction, i.e., $\pm\hat{\mathbf{z}}$. Therefore, $\mathbf{j}_{sCPGE}$ circulates around the beam center with an amplitude proportional to the length of $\mathbf{q}$ and a sign determined by the photon helicity, as observed experimentally. The measured photocurrent magnitude can be related to $\mathbf{j}_{sCPGE}$ by a geometric factor associated with the electrode positions, and is captured by a simple electrostatic model (see supplementary note 2). When scanning the laser spot position perpendicular to the two electrodes (Fig. 3a), $J_C$ collected by the electrode pair reflects the antisymmetric dependence on spatial coordinates of $\mathbf{j}_{sCPGE}$, and can be well fitted. The dependence of $J_C$ on the Gaussian beam width when the laser beam position is fixed (Fig. 3b) is also reproduced by our model, indicating that the phenomenological expression is consistent with the experimental data.

To develop a general microscopic description for the observed s-PGE, we studied the semiclassical quantum density matrix $\rho(\mathbf{r}, \mathbf{k}, \mathbf{t})$ to first order in spatial field gradients and second order in the electric field within a nonlinear susceptibility framework[34] (see Supplementary Information, Note 3). The general quantum kinetic equation[35] is obtained from the equation of motion for the Wigner transformation for $\rho$, which includes the spatial inhomogeneity of $\rho$ through the electric field driving term. Analogous to the injection current[36] in a homogenous system, the derived steady state response functions of $\mathbf{j}_{sPGE}$ consist of $\rho^{(2)}$ quadratic in $\mathbf{E}$ and linear in $\mathbf{q}$, and the band diagonal velocity, $\mathbf{v}_{nn} = \frac{\partial \varepsilon_n(\mathbf{k})}{\partial \mathbf{k}}$ with $\varepsilon_n(\mathbf{k})$ being the energy of band $n$ at Bloch momentum $\mathbf{k}$. We assume that the measured $\mathbf{j}_{sPGE}$ is dominated by electronic interband transitions due to the high photon energy of the excitation beam (~1.65eV). The two terms that control the conductivity tensor $\beta$ for $\mathbf{j}_{sCPGE}$ in Eqn (2) are (see Supplementary Information, Note 4);

$$\beta_{ilj,1} = \sum_{\mathbf{k},n,m} \frac{ie^3}{2\hbar^2}(\Gamma_{nm}(\omega) + \Gamma_{mn}(-\omega))(f_0(\varepsilon_m) - f_0(\varepsilon_n))\Omega^j_{nm}(k)(v^l_{nn}v^i_{nn}\tau^2_{nn} - v^l_{mm}v^i_{mm}\tau^2_{mm}) \quad (3)$$

$$\beta_{ilj,2} = \sum_{\mathbf{k},n,m} \frac{e^3}{8\hbar}(\Gamma^2_{nm}(\omega) + \Gamma^2_{mn}(-\omega))(f_0(\varepsilon_m) - f_0(\varepsilon_n))\Omega^j_{nm}(k)(v^l_{nn} + v^l_{mm})(v^i_{nn}\tau_{nn} - v^i_{mm}\tau_{mm}) \quad (4)$$

where $f_0(\varepsilon_n(\mathbf{k}))$ is the Fermi-Dirac distribution, $\tau_{nn}(\mathbf{k})$ is the relaxation time of excited carriers in band $n$, $\Gamma_{nm}(\mathbf{k}, \omega) = \frac{1}{\hbar\omega + \varepsilon_n - \varepsilon_m - \frac{i\hbar}{\tau_{nm}}}$, and $\Omega^i_{nm}(\mathbf{k}) = -i(R^j_{nm}R^k_{mn} - R^k_{nm}R^j_{mn})$ is derived from interband matrix elements of the non-abelian Berry connection, $\mathbf{R}_{nm}(\mathbf{k})$.

This quantity transforms like the Berry curvature: $\Omega^i_{nm}(\boldsymbol{k}) = \Omega^i_{nm}(-\boldsymbol{k})$ under inversion symmetry and $\Omega^i_{nm}(\boldsymbol{k}) = -\Omega^i_{nm}(-\boldsymbol{k})$ under time reversal symmetry so that it is allowed only if time reversal or inversion symmetry are broken. These are precisely the conditions which permit Weyl nodes in the spectrum. However, the observations reported here are responses to excitations well above the Lifshitz energy where may not directly access the topological character of the low energy excitations. Since these response functions carry one higher order of the band diagonal velocities in comparison to conventional injection current, $\mathbf{j}_{sPGE}$ magnitude will be more sensitive to band dispersion. Therefore, the expressions explain why s-CPGE does not exist in the 1T' phase of MoTe$_2$ but arise only after a temperature or doping induced phase transition to the broken inversion phase.

Unlike conventional injection current, momentum space asymmetry in the electron scattering rate is crucial for the existence of s-CPGE since only the antisymmetric contribution to the relaxation time, $\tau_{nm}(-\boldsymbol{k})^{(a)} = -\tau_{nm}(\boldsymbol{k})^{(a)}$ can give rise to nonzero $\beta_1$ or $\beta_2$. In general, the scattering probability function follows the crystal symmetry[37], so an antisymmetric modulation of the relaxation time of **k** is allowed only in the broken inversion phase in these materials. Furthermore, when large spin-orbit coupling is present, spin-dependent skew scattering[38-40] of positive and negative **k** states occur with different probabilities, $(i.e.\ W_{kk'} \neq W_{k'k})$ and would augment an isotropic scattering rate by an antisymmetric contribution which is the main contribution to $\tau_{nm}(\boldsymbol{k})^{(a)}$.

This microscopic description of the s-CPGE response requires controlling the k-space distribution of excited electrons using optical field gradients, in contrast to conventional CPGE which uses only the polarization of a spatially uniform optical field. The effect is illustrated by a transition between the valence and conduction bands under Gaussian beam excitation, shown in Fig. 3c. The bands are colored by the difference of excitation probability (nonequilibrium electron population distribution) under RCP and LCP light, derived from the density matrix formalism. Contrary to the conventional CPGE where the electrons would be excited following the intrinsic distribution of $\Omega_{nm}^{i}(\boldsymbol{k})$, for s-CPGE (Fig. 3c inset), the excitation probability has an asymmetric component, which changes sign when the local **q** is reversed (i.e. on the opposite side of the Gaussian beam), showing that the interaction of the bands with the optical field can be controlled by the beam profile.

Overall, the s-CPGE response in MoTe$_2$/Mo$_{0.9}$W$_{0.1}$Te$_2$ expressed in terms of $\Omega_{nm}^{i}(\boldsymbol{k})$ is allowed by the broken inversion symmetry and is related to a large spin-orbit interactions (SOI) in Weyl semimetals[41]. This naturally raises a question about the effect of band crossings in a WSM on s-CPGE phenomena. To study the possibility of s-CPGE for low frequency excitations below the Lifshitz energy near the Weyl cone, we calculated our response functions using a minimal model describing a phase transition from a Dirac to a Weyl semimetal. A $4 \times 4$ Hamiltonian describing a three–dimensional Dirac semimetal is adopted[42],

$$H_\Gamma(\boldsymbol{k}) = \varepsilon_0(\boldsymbol{k}) + M(\boldsymbol{k})\mathbb{1}\otimes\tau_z + Ak_x\sigma_z\otimes\tau_x - Ak_y\mathbb{1}\otimes\tau_y \tag{5}$$

where $\varepsilon_0(\boldsymbol{k}) = C_0 + C_1 k_z^2 + C_2(k_x^2 + k_y^2)$, $k_\pm = k_x \pm ik_y$ and $M(\boldsymbol{k}) = M_0 - M_1 k_z^2 - M_2(k_x^2 + k_y^2)$. Degenerate bands form two Dirac points $(0,0, \pm\sqrt{\frac{M_0}{M_1}})$ along the $\hat{k}_z$ axis (Fig. 4a) and because of time reversal and inversion symmetry, $\Omega_{nm}^{ij}(\boldsymbol{k})$ is zero, and all relevant terms in s-CPGE exactly vanish. Upon adding a small inversion breaking term controlled by the parameter $L_0$, the system transforms from a Dirac semimetal to a Weyl semimetal

$$H'_\Gamma(\boldsymbol{k}) = H_\Gamma(\boldsymbol{k}) + L_0 k_z \sigma_z \otimes \tau_z \tag{6}$$

This WSM has four Weyl points separated along the $\hat{\boldsymbol{k}}_z$ axis (Fig. 4b). $\Omega^x_{23}(\boldsymbol{k})$ plotted in Fig. 4c now has hot spots at the Weyl points with signs related to the Chern numbers +1 and -1, similar to the Berry curvature. The nonvanishing $\Omega^i_{nm}(\boldsymbol{k})$ produces a nonzero $\boldsymbol{j}_{sCPGE}$ and the transverse part of the conductivity, $\sigma_{zyx}$, as a function of frequency ω is shown in Fig. 4d. Numerical analysis shows that $\sigma(\omega)$ scales as $\frac{\alpha\omega + \alpha'\omega^2}{1+\beta\omega^3+\beta'\omega^4}$ (where $\alpha, \alpha', \beta, \beta'$ are fitting parameters). When the Fermi level is above the Lifshitz energy (but still within the band width defined by the energy cutoff), σ grows linearly with slope $\alpha$ due to Pauli blocking at small frequencies and reaches a maximum at $\omega_p = \frac{1}{(2\beta)^3}$ followed by $\omega^{-2}$ scaling at the high frequency tail. This behavior is different from conventional CPGE where instead σ(ω) scales as $\omega^{-1}$ at high frequencies (Fig. S4). However, when the Fermi level is less than the Lifshitz energy, $\sigma_{zyx}$ changes its sign at a certain frequency determined by the chemical potential, which is related to band crossing in Weyl semimetals. Upon increasing the inversion breaking parameter $L_0$, the initial slope $\alpha$ gradually increases, leading to a stronger s-CPGE (Fig 4d inset). Overall, since s-CPGE shows both a different scaling behavior in comparison to CPGE at high frequency and is sensitive to detailed band parameters and topology at low frequency, it may be a very useful spectroscopic probe of these materials. More generally it can be applied to control photogalvanic response via the patterning of light intensity distribution and polarization.

In conclusion, a strong spatially dispersive CPGE with photon helicity dependent circulating photocurrent is observed in type-II WSMs $MoTe_2$ and $Mo_{0.9}W_{0.1}Te_2$. The newly derived nonlinear susceptibilities encode the effects of spatially inhomogeneous field excitation and explain the existence of s-CPGE in WSMs. In this framework, these effects are attributed to the inversion symmetry breaking and asymmetric carrier excitation in momentum space due to optical field gradients. Our work also demonstrates that precisely tailored photon spin-dependent optoelectronic responses can be engineered in these systems by shaping and patterning optical field profiles, which can greatly enhance the applications of topological materials over a broad spectral range.

**Methods**

**Growth of single crystals**: Large, well-formed, ribbon-like single crystals of MoTe$_2$ and Mo$_{0.9}$W$_{0.1}$Te$_2$ alloy were grown by chemical vaper transport (CVT) with iodine (I) as the carrier gas. Stoichiometric amounts of tungsten (W) powder (99.9%, Sigma-Aldrich), molybdenum (Mo) powder (99.95%, Sigma-Aldrich) and tellurium (Te) powder (99.95%, Sigma-Aldrich) with a total weight of 500 mg, plus an extra 35 mg of I as the transport gas were sealed in an evacuated 20 cm long quartz tube under vacuum at $10^{-6}$ Torr. The quartz tube was placed in a three-zone furnace. Firstly, the reaction zone was maintained at 850 °C for 30 h with the growth zone at 900 °C in order to prevent the transport of the product and a complete reaction; then the reaction zone was heated to 1070 °C and held for 7 days with the growth zone at 950 °C. Finally, the furnace was naturally cooled down to room temperature and the single crystals were collected in the growth zone. Residual I was cleaned using acetone before measurement.

**Device fabrication:** Devices were fabricated on exfoliated MoTe$_2$ or Mo$_{0.9}$W$_{0.1}$Te$_2$ flakes with thickness ranging from 100~300 nm and typical dimensions being 20 um ×20 um assembled on SiO$_2$/Si substrates. Electrodes were defined by electron beam lithography followed by physical vapor deposition of 300 nm Ti/100 nm Au film.

**Photocurrent measurements:** The excitation source was provided from a wavelength tunable Ti-Sapphire pulsed laser in the 680-1020 nm. The laser was focused to a near perfect Gaussian spot by a 60X objective and the full width at half maximum (FWHM) of the spot was controlled in the range of ~2-20 µm FWHM, with total power in the 1-10 mW range. Quarter wave plate (QWP) mounted on a motorized precision rotation stage driven by a servo motor (Thorlabs) was used to vary the angle continuously from 0-360° to obtain different laser polarizations. The laser polarization on the sample plane was analyzed carefully to ensure accuracy. The power difference between the left and right circularly polarized light was measured by the power meter to be less than 1%, and the extinction ratio of linearly polarized light was ensured to be larger than 1000:1. When scanning the light beam over the sample using piezoelectric stages, the spatial coordinates were recorded with an accuracy of ~200 nm. Photocurrents were recorded using a current preamplifier (DL instruments model 1211) for which the voltage bias was sourced and the output signal from the preamplifier (photocurrent was converted to an amplified voltage signal) recorded continuously (~10 data points per second) by the PCI card (National Instrument, NI PCI-6281). The time constant of the preamplifier was chosen in the range of 100-300 ms. The quarter wave plate was rotated at the rate of ∼7°/sec using a motorized precision rotation stage with a servo motor[27].

**Figure Captions:**

**Figure 1**. **Polarization-dependent photocurrent measurements on 1T′ (300 K) and T$_d$ (Weyl, 77 K) phases of MoTe$_2$.** (**a, b**) Crystal structures of 1T′ (a) and T$_d$ phase of MoTe$_2$ (b). Yellow (purple) spheres represent Te (Mo) atoms. (**c**) Schematic of the polarization-dependent photocurrent measurement setup. In all our experiments, a Gaussian laser beam propagating along the **z** axis was focused by a microscope objective (60X) incident normally onto the sample (**x-y** plane; spot size ≈2um), with the **z** axis parallel to the crystallographic *c* axis of MoTe$_2$. (**d-g**) Photocurrent plotted as a function of quarter waveplate fast axis rotation angle $\varphi$ at two spatial locations and temperatures: (**d**) spot **a** at 300K; (**e**) spot **b** at 300K; (**f**) spot **a** at 77K; (**g**) spot **b** at 77K. Black dots are the experimental data and red solid lines are the fits to Eqn (1).

**Figure 2**. **Measurement of circulating current in the T$_d$ (Weyl) phase of Mo$_{0.9}$W$_{0.1}$Te$_2$ at room temperature under circularly polarized optical excitation**. (**a**) Optical image of the multi-electrode Mo$_{0.9}$W$_{0.1}$Te$_2$ device (**x-y** plane). The five electrodes are labeled a-e, and red and blue arrows indicate the circulating direction of CPGE current under left and right circularly polarized light illumination (spot size ≈ 2um) respectively. (**b-f**) Photocurrents measured between each of the nearest electrode pairs and plotted as functions of the fast axis rotation angle, $\varphi$, of the quarter waveplate. Plots correspond to measurements performed between electrodes (**b**) a→b, (**c**) b→c, (**d**) c→d, (**e**) d→e, and (**f**) e→a. Black dots are the experiment data, red solid lines are the fitted curves for total photocurrent (Eqn (1)), and the green solid lines represent the fitted CPGE currents, J$_C$.

The blue and red arrows represent the circulating CPGE current under RCP and LCP illumination, respectively.

**Figure 3. Spatial location and Gaussian spot size dependence of the s-CPGE current in $Mo_{0.9}W_{0.1}Te_2$ at room temperature**. (**a**) (Inset) Schematic of the spatially dependent photocurrent measurement on $Mo_{0.9}W_{0.1}Te_2$ (**x-y** plane). Laser beam (spot size ≈2um) was scanned along the **y** axis, as indicated by the green arrow. $L_e$ is the total electrode length on the sample and L is the separation between the two electrodes. (Main) CPGE current, $J_C$, as a function of the laser beam position. Black squares are the experimental data and the red solid line is the fitting curve to the electrostatic model described in Supplementary Note 2. (**b**) CPGE current, $J_C$, plotted as a function of the Gaussian beam diameter at a fixed distance $y_0$ to electrodes indicated in (**a**). Black squares are the experimental data and the red solid line is the fitting curve to the expression derived from the phenomenological model of $\mathbf{j}_{sCPGE}$. (**c**) Schematic of asymmetric interband excitation by a Gaussian beam, where the band touching points are located along the **x** axis, and light propagates in the **z** direction. (Main) Spatial intensity distribution of a Gaussian beam along the **y** axis, with $y_{center}$ being the y coordinate of the Gaussian beam center. (Inset) Local excitation patterns contributing to s-CPGE current in the momentum space at the left ($y_{left}$) and right ($y_{right}$) tails of the Gaussian beam. The color map shows the normalized difference between the excitation probability (nonequilibrium electron population) under right and left circularly polarized light illumination. A negative value (blue region) implies that in comparison to homogenous excitation, the optical field gradient results in less electrons being excited, while a positive value (red color) implies excitation of more electrons.

**Figure 4**. **Numerical results for s-CPGE current from a four band minimal model for Dirac and Weyl semimetals** (**a**) Band structure of a Dirac semimetal[42]. (**b**) Band structure of the inversion broken Weyl semimetal obtained from (**a**) by adding an inversion breaking term from Eqn. (6). (**c**) Band resolved Berry curvature plot on the $k_x = 0.01$ Å$^{-1}$ plane of the Weyl semimetal. The color map represents its intensity, and the two dashed circles mark a pair of Weyl points. (**d**) Plots of the transverse s-CPGE conductivity $\sigma_{zyx}$ of the Weyl semimetal as a function of optical frequency, ω at different Fermi energies and fittings to the scaling function $\frac{\alpha\omega + \alpha'\omega^2}{1+\beta\omega^3+\beta'\omega^4}$. Inset shows the dependence of the fitting parameter $\alpha$ on the inversion breaking parameter $L_0$, at the Fermi energy $\mu = 0.3$ eV.

**Tables**

| MoTe$_2$ | spot a | | spot b | |
|---|---|---|---|---|
| | $J_L$ (nA) | $J_C$ (nA) | $J_L$ (nA) | $J_C$ (nA) |
| 300K (1T') | 23 | ~0 | -30 | ~0 |
| 77K (T$_d$) | 34 | 14 | -40 | -21 |

**Table 1**: **Fitting parameters for polarization dependent photocurrent data measured on MoTe$_2$ at two different spatial locations (a and b) at two temperatures shown in Fig. 1(d-g) using Eqn (1). $J_L$ and $J_C$ are the LPGE and CPGE components, respectively.**

**Acknowledgements:** RA acknowledges the support from the Office of Naval Research MURI (grant #N00014-17-1-2661) and the US Army Research Office (grant # W911NF-17-1-0436). Work by EJM and ZA was supported by the Department of Energy (grant# DE FG02 84ER45118). The crystal growth effort (PY and ZL) was supported by the Singapore National Research Foundation under NRF RF Award No. NRF-RF2013-08 and Tier 2 MOE2016-T2-2-153. RA, EJM, CLK and AMR acknowledge the support from the Center of Excellence for Materials Research and Innovation Seed Grant and the MRSEC iSuperSEED Supplement.

**Competing financial interests:** The authors declare no competing financial interests.

**Figure 1:**

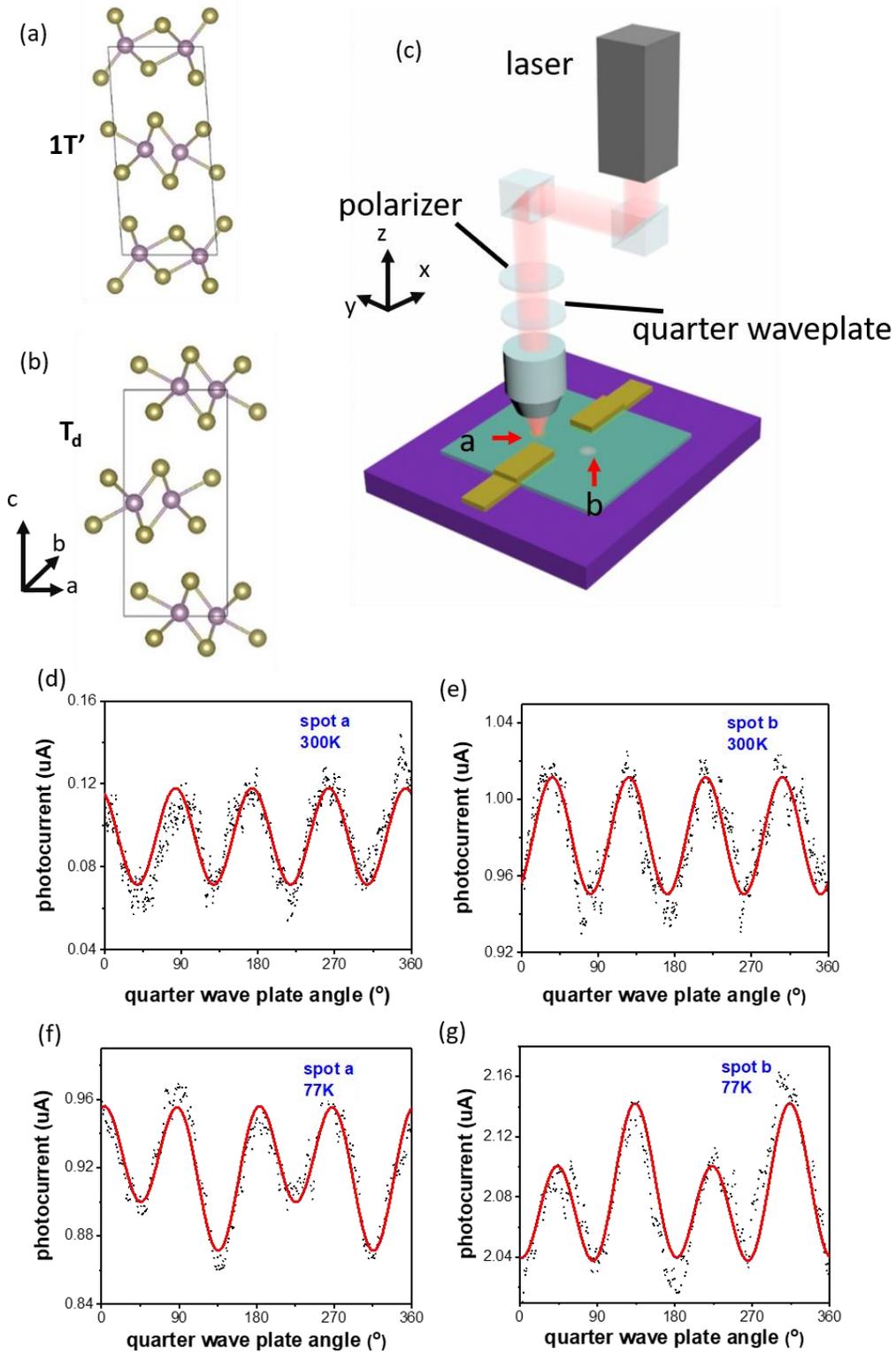

**Figure 2:**

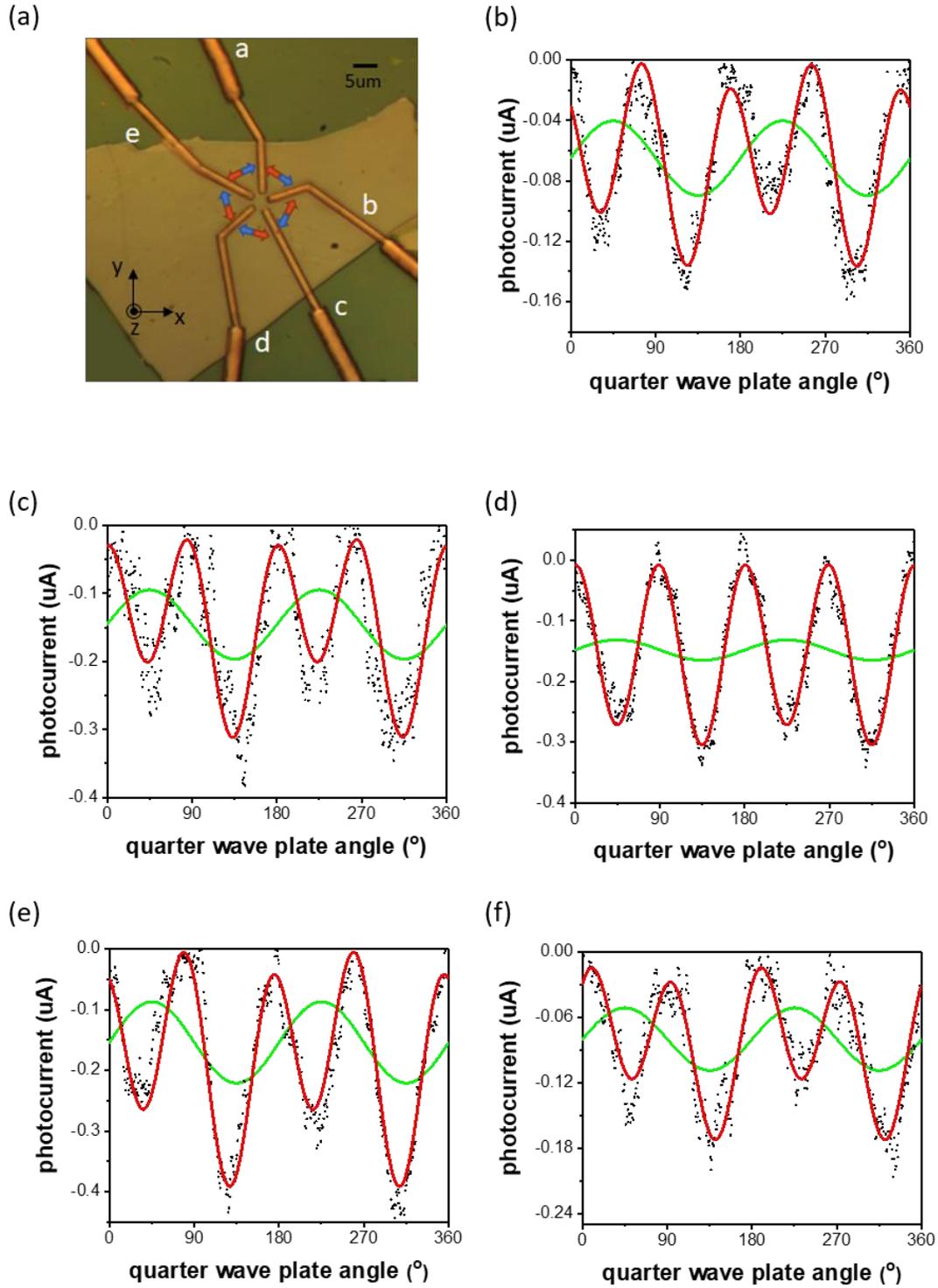

**Figure 3:**

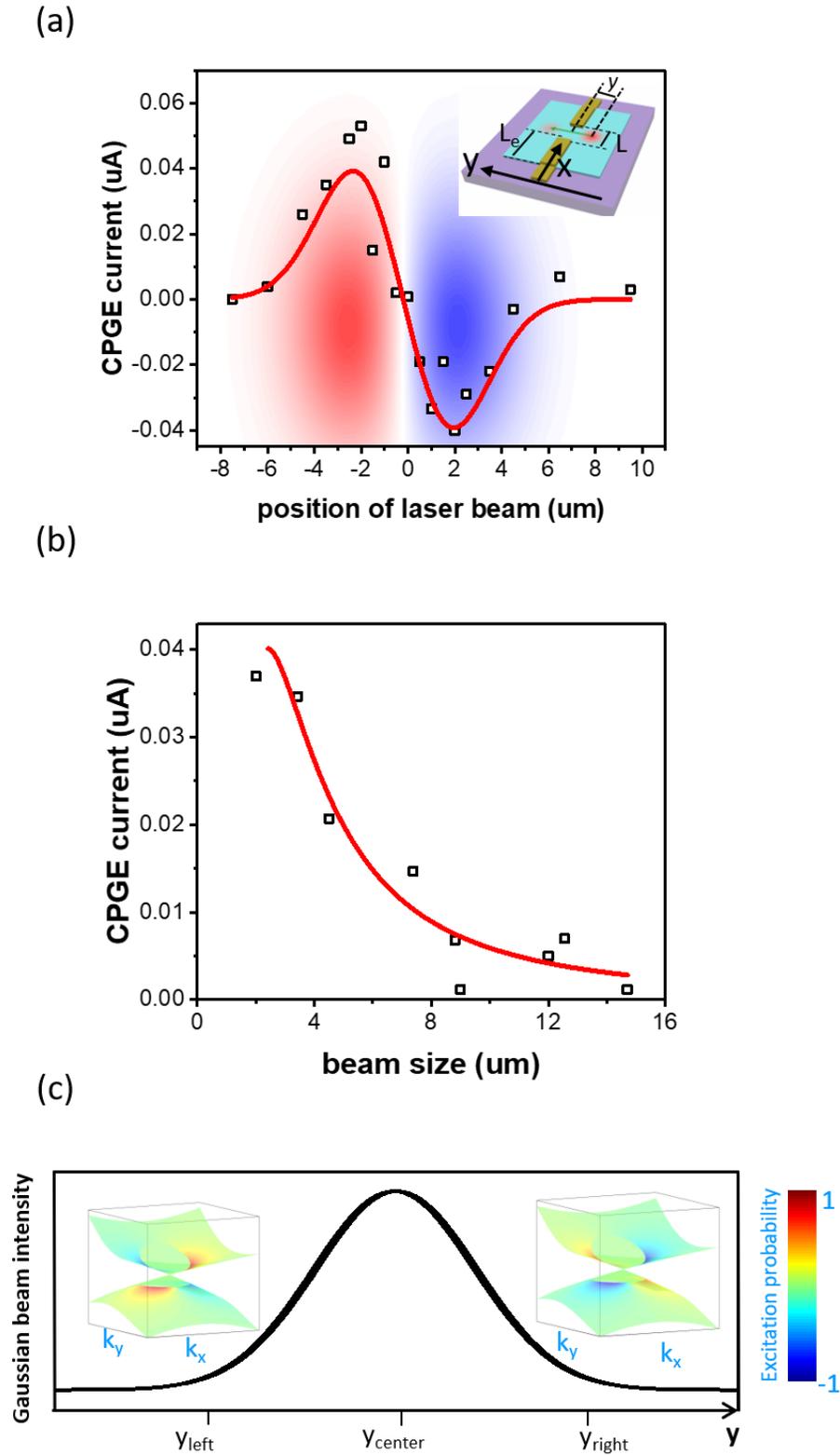

**Figure 4:**

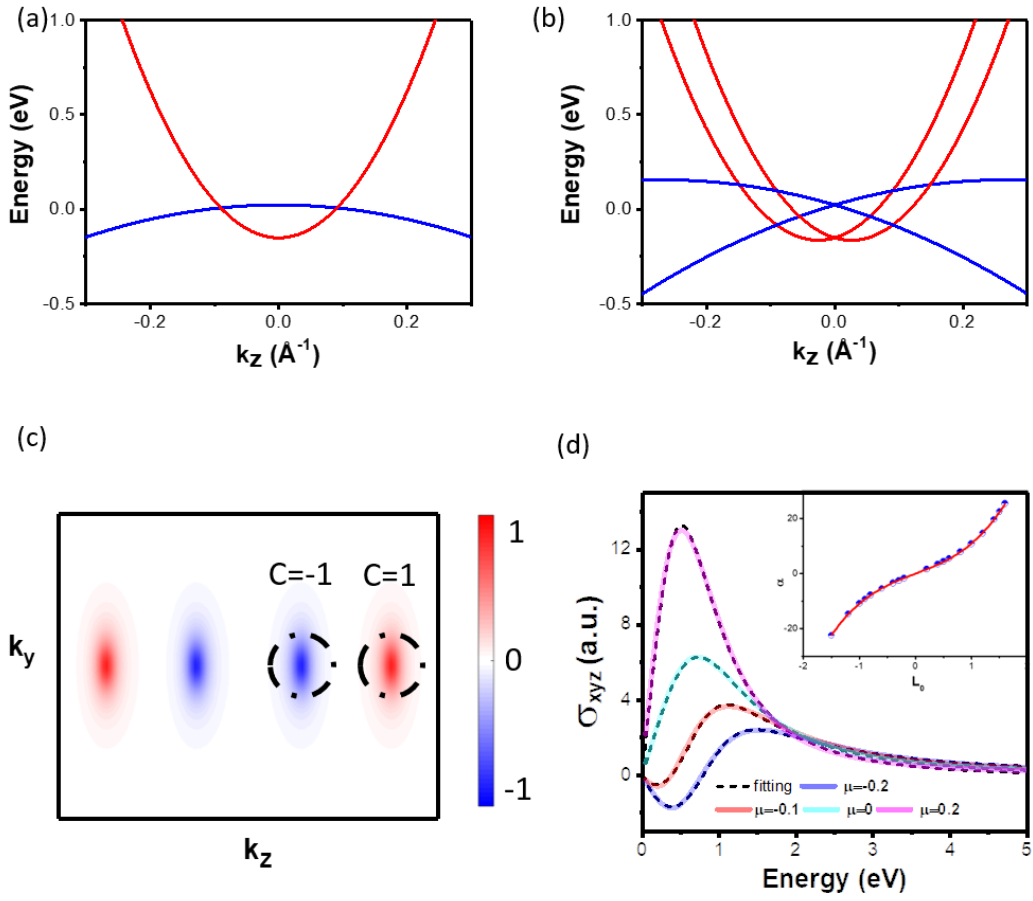